# Robust Linear Design for Flight Control Systems with Operational Constraints

Marcel Menner[1]
*Aurora Flight Sciences, Cambridge, MA, 02142, USA*

Eugene Lavretsky[2]
*The Boeing Company, Huntington Beach, CA, 92647, USA*

This paper presents a systematic approach for designing robust linear proportional-integral (PI) servo-controllers that effectively manage control input and output constraints in flight control systems. The control design leverages the Nagumo Theorem and the Comparison Lemma to prove constraint satisfaction, while employing min-norm optimal controllers in a manner akin to Control Barrier Functions. This results in a continuous piecewise-linear state feedback policy that maintains the analyzability of the closed-loop system through the principles of linear systems theory. Additionally, we derive multi-input multi-output (MIMO) robustness margins, demonstrating that our approach enables robust tracking of external commands even in the presence of operational constraints. Moreover, the proposed control design offers a systematic approach for anti-windup protection. Through flight control trade studies, we illustrate the applicability of the proposed framework to real-world safety-critical aircraft control scenarios. Notably, MIMO margin analysis with active constraints reveals that our method preserves gain and phase margins comparable to those of the unconstrained case, in contrast to controllers that rely on hard saturation heuristics, which suffer significant performance degradation under active constraints. Simulation results using a nonlinear six-degree-of-freedom rigid body aircraft model further validate the effectiveness of our method in achieving constraint satisfaction, robustness, and effective anti-windup protection.

## I. Introduction

Robustness and control margin analysis are central to both airworthiness evaluation and the development of flight control systems (FCS), ensuring safety while maintaining practical applicability. Robustness ensures that an aircraft can handle unforeseen situations that deviate from nominal flight conditions. However, the effectiveness of these analyses relies on the validity of linear system theory [1], [2], which assumes that linear models accurately represent the aircraft and its FCS. While linear representations are widely accepted within nominal operating regions, their validity can diminish under conditions such as actuator saturation or when system outputs exceed operational limits. To address these challenges, heuristic modifications like anti-windup protection are often employed to mitigate undesirable transients during saturation events [3]. However, as aircraft configurations become more complex, particularly those designed for vertical takeoff and landing, reliance on heuristics may not provide sufficient assurance of airworthiness.

This paper builds upon our findings presented in [4], where we proposed a systematic approach to servo-control design for multi-input-multi-output (MIMO) linear time-invariant (LTI) systems with operational constraints. This paper presents a proportional-integral (PI) servo-control design for FCS that incorporates min/max (box) operational constraints applied component-wise to both control inputs and limited outputs. The proposed control augmentation is based on the Nagumo Theorem [5] (English translation [6]), the Comparison Lemma [7], and min-norm optimal control design principles [8]. The Nagumo Theorem ensures forward invariance of closed-loop system trajectories, while the Comparison Lemma aids in deriving operational dynamics constraints enforceable by the LTI system. We embed these linear dynamics constraints into a Quadratic Program (QP) formulation, allowing for the enforcement of designated operational constraints. This feature is particularly beneficial for controllers utilizing integral feedback, as it systematically addresses integrator windup. By employing a formal design method to enforce box constraints through feedback, we move away from the ad hoc logic often used in flight control applications. This method enables the derivation of an explicit analytical solution to the QP [9]. The resulting control design yields a continuous piecewise-linear state feedback policy, for which system stability and robustness metrics can be computed using linear systems theory. Fig. 1 illustrates the overall control block diagram, which includes the baseline PI servo-controller with the min-norm optimal state feedback augmentation. In this paper, we derive MIMO robustness margins specifically for scenarios involving active control input and output constraints. Our findings provide valuable insights into potential solutions, including an application of anti-windup techniques that demonstrate the bounded nature of output tracking error, ensuring system stability even under constraints. We further illustrate the practical implications of our approach through the design of FCSs that account for aileron and rudder constraints, as well as roll rate and sideslip angle limitations. To validate our methodology, we present an application to a nonlinear six-degree-of-freedom simulation, showcasing the effectiveness of our systematic control design in real-world scenarios with unmodeled dynamics.

---

[1] Senior Engineer, Autonomy.
[2] Senior Principal Technical Fellow, Boeing Research & Technology. AIAA Fellow.

Related work includes Control Barrier Functions (CBFs) [10], [11], [12] and model predictive control (MPC) [13], both of which have gained significant attention for their potential to enhance safety in autonomous systems. For example, MPC-based augmentation logic is often employed as a "safety filter," allowing a baseline controller to operate unconstrained as long as operational limits are not approached [14]. In contrast, this paper employs vector-valued CBFs within a min-norm optimal controller framework to derive an explicit analytical control policy rather than solving an optimization problem online. The resulting solution is given by a piecewise-linear state feedback servo-controller, which can be rigorously analyzed using control theory.

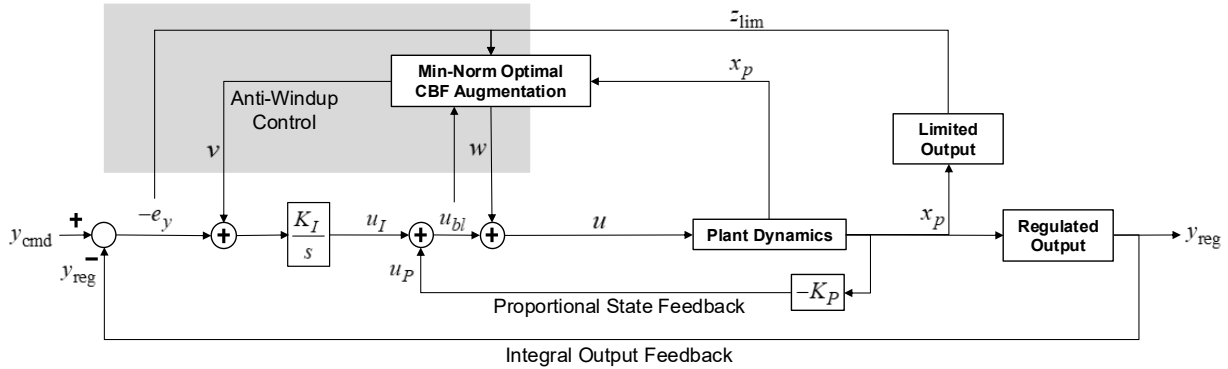

**Fig. 1. Closed-loop system block-diagram. The proportional-integral baseline controller is augmented with a control augmentation, which ensures that control input and output stay within their operational limits.**

## II. Problem Formulation

Consider the stabilizable MIMO LTI dynamical system,

$$\dot{x}_p(t) = A_p x_p(t) + B_p u(t)$$
$$y_{\text{reg}}(t) = C_{p\,\text{reg}} x_p(t) + D_{p\,\text{reg}} u(t) \quad (1)$$
$$z_{\text{lim}}(t) = C_{p\,\text{lim}} x_p(t)$$

where $x_p(t) \in R^{n_p}$ is the state vector, $u(t) \in R^m$ is the control inputs, $y_{\text{reg}}(t) \in R^m$ is the vector of regulated outputs, and $z_{\text{lim}}(t) \in R^m$ is the vector of limited outputs. Of interest is a PI servo-controller with robust tracking of external commands $y_{\text{cmd}}$, i.e., $y_{\text{reg}}(t) \to y_{\text{cmd}}$ as $t \to \infty$, while min/max operational constraints on the system control input $u(t)$ and on the selected limited outputs $z_{\text{lim}}(t)$ are satisfied for all times, component-wise,

$$u^{\min} \leq u(t) \leq u^{\max}$$
$$z_{\text{lim}}^{\min} \leq z_{\text{lim}}(t) \leq z_{\text{lim}}^{\max}. \quad (2)$$

We make the following assumptions. Note that Assumption 3 is standard in controls literature and implies that the extended pair $(A, B)$ is controllable.

*Assumption* 1. The matrix pair $(A_p, B_p)$ is stabilizable, and the state vector $x_p(t)$ is accessible for control design.

*Assumption* 2. The dynamical system (1) has a well-defined vector relative degree, see Section III-A for details

*Assumption* 3. The regulated output $y_{reg}(t)$ of the dynamical system in (1) has no transmission zeros at the origin.

## A. Controller Formulation and Problem Statement

This paper uses the control formulation developed in [4] applied to PI servo-controllers for a FCS. Consider the control command,

$$u(t) = u_{bl}(t) + w(t) \tag{3}$$

with a control augmentation signal $w(t)$ and the baseline state feedback control policy,

$$u_{bl}(t) = -K_I e_{yI}(t) - K_P x_p(t), \tag{4}$$

where $K_I$ and $K_P$ represent the integral and the proportional feedback gain matrices, respectively, and an integrated output tracking error $e_{yI}(t)$. For robust tracking of external commands $y_{cmd}$ and operating in the presence of operational constraints (2), the integrated output tracking error dynamics are augmented with the anti-windup (AW) control modification term $v(t) \in R^m$,

$$\begin{aligned} \dot{e}_{yI}(t) &= e_y(t) + v(t) \\ e_y(t) &= y_{reg}(t) - y_{cmd} \end{aligned} \tag{5}$$

with the output tracking error signal $e_y(t)$. Hence this paper designs a feedback control augmentation policy $w(t)$ and $v(t)$ such that the following properties are satisfied for a safety-critical FCS dealing with limited output and/or control input constraints. The system's regulated output $y_{reg}(t)$ tracks external commands $y_{cmd}$ to the extent possible, while the control input $u(t)$ and the limited output $z_{lim}(t)$ evolve within their operational constraints (2). Further, the state $x_p(t)$ and integrated output error $e_{yI}(t)$ remain bounded for all times.

## B. Augmented System Representation

Next, the dynamical system is reformulated such that the control augmentation in [4] can be applied. First, augmenting the system dynamics in (1) with the error dynamics in (5) and the control command (3), gives the $n = (n_p + m)$ – dimensional extended system

$$\underbrace{\begin{bmatrix} \dot{e}_{yI}(t) \\ \dot{x}_p(t) \end{bmatrix}}_{\dot{x}(t)} = \underbrace{\begin{bmatrix} 0_{m \times m} & C_{p\,reg} \\ 0_{n \times m} & A_p \end{bmatrix}}_{A} \underbrace{\begin{bmatrix} e_{yI}(t) \\ x_p(t) \end{bmatrix}}_{x(t)} + \underbrace{\begin{bmatrix} I_m & D_{p\,reg} \\ 0_{n \times m} & B_p \end{bmatrix}}_{B} \left( \begin{bmatrix} -y_{cmd} \\ u_{bl}(t) \end{bmatrix} + \begin{bmatrix} v(t) \\ w(t) \end{bmatrix} \right). \tag{6}$$

Second, the limited output constraints and control input constraints are written in terms of the extended system,

$$y_{lim}(t) = \begin{bmatrix} u_{bl}(t) \\ z_{lim}(t) \end{bmatrix} = \underbrace{\begin{bmatrix} -K_I & -K_P \\ 0_{m \times m} & C_{p\,lim} \end{bmatrix}}_{C_{lim}} \begin{bmatrix} e_{yI}(t) \\ x_p(t) \end{bmatrix} = C_{lim} x(t), \quad y^{max} = \begin{bmatrix} u^{max} \\ z_{lim}^{max} \end{bmatrix}, \quad y^{min} = \begin{bmatrix} u^{min} \\ z_{lim}^{min} \end{bmatrix}, \tag{7}$$

where the constraints become

$$\begin{aligned} \text{Min Constraints} &: h_{min}(x(t)) = y_{lim}^{min} - C_{lim} x(t) \leq 0 \\ \text{Max Constraints} &: h_{max}(x(t)) = C_{lim} x(t) - y_{lim}^{max} \leq 0. \end{aligned} \tag{8}$$

Importantly, in [4], it is shown that limiting the baseline control policy $u_{bl}(t)$ in (7) implies that the total control command $u(t)$ evolves within the same bounds.

### III. Constrained Quadratic Program for Servo-Control Augmentation Design

The control design is related to the method of CBFs, which use the same three theoretical pillars: The Nagumo Theorem, the Comparison Lemma, and a min-norm optimal control design concept. However, the framework in this paper uses a min-norm optimal control design only for design and no optimization problem needs to be solved online during system operation. Nonetheless, in order to acknowledge historical precedence and originality of the CBF design, the developed method shall be often referred to as the "CBF augmentation." In this section, a QP is formulated using a modification of the operational limits (2) based on the Comparison Lemma [7]. This QP uniquely defines the CBF augmentation $v(t)$ and $w(t)$ to render the system (1) forward invariant with respect to the operational limits. Motivated by the min-norm controller design method [8], consider the following QP [9],

$$\text{Cost}: J(v,w) = \left( \begin{bmatrix} v \\ w \end{bmatrix}^T R_\pi \begin{bmatrix} v \\ w \end{bmatrix} \right) \to \min_\pi$$

$$\text{Constraints}: h(x) = \begin{bmatrix} h_{\min}(x) \\ h_{\max}(x) \end{bmatrix} = \begin{bmatrix} y_{\lim}^{\min} - C_{\lim} x \\ C_{\lim} x - y_{\lim}^{\max} \end{bmatrix} \leq 0. \tag{9}$$

In order to solve QP (9), the constraints needs to be written explicitly in terms of the control augmentation signals. This can be accomplished by differentiating the constraints in (9) along the closed-loop system trajectories and leveraging the Comparison Lemma [7], i.e., the original constraints $h(x)$ are replaced by their respective differentiated versions, which is discussed next.

#### A. Modified Constraints for Controller Design

Next, the output and input constraints (2) are modified to enable controller integration. Suppose that the relative degree of $y_{\lim,i}(t)$ is $r_i \geq 1$ for $i = 1, \ldots, 2m$, where relative degree refers to the number of time a limited output needs to be differentiated for a control augmentation signal to appear explicitly. Further, suppose that the limited output of the system $y_{\lim}$ has a vector relative degree $r = (r_1 \ \ldots \ r_{2m})$ [15], i.e.,

$$\left[ \left\| \nabla_{\begin{bmatrix} v \\ w \end{bmatrix}} \left( y_i^{(k)} \right) \right\| = 0, \forall 1 \leq k \leq r_i - 1 \right] \wedge \left[ \left\| \nabla_{\begin{bmatrix} v \\ w \end{bmatrix}} \left( y_i^{(r_i)} \right) \right\| \neq 0 \right]. \tag{10}$$

Thus, for $y_{\lim}$ with the vector relative degree $r$, the control-to-output sensitivity matrix $H_u \in R^{2m \times 2m}$ must be nonsingular,

$$H_u = \begin{bmatrix} (C_{\lim})_1 A_{cl}^{r_1 - 1} B \\ \vdots \\ (C_{\lim})_{2m} A_{cl}^{r_{2m} - 1} B \end{bmatrix}, \quad \det(H_u) \neq 0. \tag{11}$$

The proposed control design uses the vector relative degree definition and (11) to differentiate the operational constraints $r-$ times (component-wise) as follows. Consider a stable polynomial of order $r_i$, with the real roots $\{\lambda_{ij}\}_{j=1,\ldots,r_i}$ with $\lambda_{ij} < 0$,

$$\phi_i(s) = \prod_{j=1}^{r_i}(s - \lambda_{ij}) = \sum_{j=0}^{r_i} c_{ij} s^j, \tag{12}$$

where $c_{ij}$ denotes the $j^{th}$ coefficient of the $i^{th}$ polynomial, and by the definition, $c_{i0} > 0$, for every $i = 1, \ldots, 2m$. The stable polynomials in (12) are used to modify the constraints in (8), resulting in

$$Y_{\lim}(t) = \underbrace{\begin{bmatrix} \phi_1(s) & \cdots & 0 \\ \vdots & \ddots & \vdots \\ 0 & \cdots & \phi_{2m}(s) \end{bmatrix}}_{\Phi(s)} y_{\lim} = \Phi(s) y_{\lim}(t)$$

$$H(x(t), u_{bl}(t), v(t), w(t)) = \begin{bmatrix} H_{\min}(x(t), u_{bl}(t), v(t), w(t)) \\ H_{\max}(x(t), u_{bl}(t), v(t), w(t)) \end{bmatrix} \tag{13}$$

$$= \begin{bmatrix} \Phi(s) h_{\min}(x(t)) \\ \Phi(s) h_{\max}(x(t)) \end{bmatrix} = \begin{bmatrix} \Phi(s)\left(y_{\lim}^{\min} - y_{\lim}(t)\right) \\ \Phi(s)\left(y_{\lim}(t) - y_{\lim}^{\max}\right) \end{bmatrix} \leq 0,$$

where $(h_{\min}(x(t)), h_{\max}(x(t)))$ are the original output constraint functions from (8), with stable polynomials $\phi_i(s)$ (12), treated as differentiation operators with respect to $s$. Then, Lemma 1 in [4] shows that the modified output $Y_{\lim}(t)$ and modified constraints can be written as

$$Y_{\lim}(t) = H_x x(t) + H_u \left( \begin{bmatrix} -y_{cmd} \\ u_{bl}(t) \end{bmatrix} + \begin{bmatrix} v(t) \\ w(t) \end{bmatrix} \right)$$

$$\begin{bmatrix} H_{\min}(x(t), u_{bl}(t), v(t), w(t)) \\ H_{\max}(x(t), u_{bl}(t), v(t), w(t)) \end{bmatrix} = \begin{bmatrix} -Y_{\lim}(t) + \alpha_\pi y_{\lim}^{\min} \\ Y_{\lim}(t) - \alpha_\pi y_{\lim}^{\max} \end{bmatrix} = \begin{bmatrix} -H_u \\ H_u \end{bmatrix} \begin{bmatrix} v(t) \\ w(t) \end{bmatrix} + \begin{bmatrix} \Delta H_{\min}(x(t), u_{bl}(t)) \\ \Delta H_{\max}(x(t), u_{bl}(t)) \end{bmatrix} \leq 0 \tag{14}$$

with

$$\Delta H_{\min}(x(t), u_{bl}(t)) = -H_x x(t) - H_u \begin{bmatrix} -y_{cmd} \\ u_{bl}(t) \end{bmatrix} + \alpha_\pi y_{\lim}^{\min}$$

$$\Delta H_{\max}(x(t), u_{bl}(t)) = H_x x(t) + H_u \begin{bmatrix} -y_{cmd} \\ u_{bl}(t) \end{bmatrix} - \alpha_\pi y_{\lim}^{\max} \tag{15}$$

and

$$H_x = \begin{bmatrix} (C_{\lim})_1 \prod_{j=1}^{r_1}(A - \lambda_{1j} I_n) \\ \vdots \\ (C_{\lim})_{2m} \prod_{j=1}^{r_{2m}}(A - \lambda_{2mj} I_n) \end{bmatrix}, \quad H_u = \begin{bmatrix} (C_{\lim})_1 A^{r_1 - 1} B \\ (C_{\lim})_2 A^{r_2 - 1} B \\ \vdots \\ (C_{\lim})_{2m} A^{r_{2m} - 1} B \end{bmatrix} \tag{16}$$

are the system state and control sensitivity matrices respectively, and $\alpha_\pi \in R^{2m \times 2m}$ is a diagonal matrix, with its positive diagonal elements defined as the zero-order coefficients of the $i^{th}$ polynomial $\phi_i(s)$ (12),

$$\alpha_\pi = \begin{bmatrix} c_{10} & \cdots & 0 \\ \vdots & \ddots & \vdots \\ 0 & \cdots & c_{2m0} \end{bmatrix}, \quad c_{i0} = \prod_{j=1}^{r_i}(-\lambda_{ij}) > 0, \quad \forall i = 1, \ldots, 2m. \tag{17}$$

Further, Lemma 2 in [4] showed that if a dynamical system is invariant forward in time with respect to the modified constraints (14), then the dynamical system in (1) is also invariant forward in time with respect to the original constraints (8). Hence, designing the control augmentation to satisfy (14) will enforce (8) for all times. For the PI servo-controller in (4), $H_u$ becomes

$$H_u = \begin{bmatrix} (C_{\lim})_1 A^{r_1-1} B \\ (C_{\lim})_2 A^{r_2-1} B \\ \vdots \\ (C_{\lim})_{2m} A^{r_{2m}-1} B \end{bmatrix} = \begin{bmatrix} -K_I & -K_I D_{p\,\text{reg}} - K_P B_p \\ 0_{m \times m} & H_w \end{bmatrix}, \quad H_w = \begin{bmatrix} (C_{p\lim})_1 A_p^{r_{m+1}-1} \\ \vdots \\ (C_{p\lim})_m A_p^{r_{2m}-1} \end{bmatrix} B_p. \tag{18}$$

**Remark 1**. Replacing inequality constraints such as (8) with their modified versions (14) that contain linear combinations of time derivatives of the original constraints and the original constraints is predicated on the Comparison Lemma [7], which in turn provides sufficient conditions to satisfy the original constraints (8) while enforcing their modified versions.

**Remark 2**. Control input constraints in (7) are, by construction, relative degree one, i.e., the control input constraint needs to be differentiated once for the control augmentation $v(t), w(t)$ to appear explicitly,

$$\begin{aligned} \dot{u}_{\text{bl}}(t) &= -K_I \left( e_y(t) + v(t) \right) - K_P \left( A_p x_p(t) + B_p \left( u_{\text{bl}}(t) + w(t) \right) \right) \\ &= -K_I \left( -y_{\text{cmd}} + v(t) \right) - \left( K_I C_{p\,\text{reg}} + K_P A_p \right) x_p(t) - \left( K_I D_{p\,\text{reg}} + K_P B_p \right) \left( u_{\text{bl}}(t) + w(t) \right). \end{aligned} \tag{19}$$

## B. QP with Modified Operations Constraints and Control Augmentation Solution

Since the control augmentation $v(t)$ and $w(t)$ appears explicitly in (14), the modified inequality constraints in (14) can directly be used for control design. To this end, consider the following QP,

$$\text{Cost}: J(v, w) = \left( \begin{bmatrix} v \\ w \end{bmatrix}^T R_\pi \begin{bmatrix} v \\ w \end{bmatrix} \right) \to \min_\pi$$

$$\text{Constraints}: H(x, u_{\text{bl}}, v, w) = \begin{bmatrix} H_{\min}(x, u_{\text{bl}}, v, w) \\ H_{\max}(x, u_{\text{bl}}, v, w) \end{bmatrix} = \begin{bmatrix} -H_u \\ H_u \end{bmatrix} \begin{bmatrix} v \\ w \end{bmatrix} + \begin{bmatrix} \Delta H_{\min}(x, u_{\text{bl}}) \\ \Delta H_{\max}(x, u_{\text{bl}}) \end{bmatrix} \leq 0 \tag{20}$$

with $R_\pi = H_u^T H_u$. Theorem 2 in [4] can now be used to show that the explicit piece-wise linear control augmentation, given by

$$\begin{bmatrix} v(t) \\ w(t) \end{bmatrix} = H_u^{-1} \left( \max \left( 0_{2m \times 1}, \Delta H_{\min}(x(t), u_{\text{bl}}(t)) \right) - \max \left( 0_{2m \times 1}, \Delta H_{\max}(x(t), u_{\text{bl}}(t)) \right) \right)$$

$$= H_u^{-1} \begin{cases} \left( -H_x x(t) - H_u \begin{bmatrix} -y_{\text{cmd}} \\ u_{\text{bl}}(t) \end{bmatrix} + \alpha_\pi y_{\lim}^{\min} \right), & \text{if } \left( -H_x x(t) - H_u \begin{bmatrix} -y_{\text{cmd}} \\ u_{\text{bl}}(t) \end{bmatrix} + \alpha_\pi y_{\lim}^{\min} \right) > 0 \\ -\left( H_x x(t) + H_u \begin{bmatrix} -y_{\text{cmd}} \\ u_{\text{bl}}(t) \end{bmatrix} - \alpha_\pi y_{\lim}^{\max} \right), & \text{if } \left( H_x x(t) + H_u \begin{bmatrix} -y_{\text{cmd}} \\ u_{\text{bl}}(t) \end{bmatrix} - \alpha_\pi y_{\lim}^{\max} \right) > 0 \\ 0, & \text{otherwise} \end{cases} \tag{21}$$

with $H_x$ and $H_u$ as in (15) and $\alpha_\pi$ as in (17), is optimal with respect to the modified QP (20), forward invariant with respect to both the modified constraints (14) and the original constraints (8), and the integrated output tracking error remains uniformly ultimately bounded (UUB) forward in time. In this paper, we extend the findings in [4], focusing on crucial aspects for practical application of the proposed control augmentation such as analyzability, robustness, and margin calculation as well as on application to flight controls for (nonlinear) aircraft dynamics. Due to the piece-wise linear control design, the closed-loop system remains analyzable in the presence of active constraints, i.e., linear system theory can be applied to assess stability and compute margins.

Table I summarizes the overall derived servo-control augmentation design for clarity.

| | |
|---|---|
| Open-loop LTI MIMO dynamics (1) | $\dot{x}_p = A_p x_p + B_p u$ |
| Regulated output (1) | $y_{reg} = C_{p\,reg} x_p + D_{p\,reg} u$ |
| Limited output (1) | $z_{lim} = C_{p\,lim} x_p$ |
| Output tracking error (5) | $e_y = y_{reg} - y_{cmd}$ |
| Integrator tracking error dynamics with integrator AW (5) | $\dot{e}_{yI} = e_y + v$ |
| Extended open-loop system and extended state (6) | $\begin{bmatrix} \dot{e}_{yI} \\ \dot{x}_p \end{bmatrix} = \begin{bmatrix} 0_{m \times m} & C_{p\,reg} \\ 0_{n \times m} & A_p \end{bmatrix} \begin{bmatrix} e_{yI} \\ x_p \end{bmatrix} + \begin{bmatrix} I_m & D_{p\,reg} \\ 0_{n \times m} & B_p \end{bmatrix} \left( \begin{bmatrix} -y_{cmd} \\ u_{bl} \end{bmatrix} + \begin{bmatrix} v \\ w \end{bmatrix} \right), \quad x = \begin{bmatrix} e_{yI} \\ x_p \end{bmatrix}$ |
| Control input with baseline control and augmentation (3) | $u = u_{bl} + w = -K_I e_{yI} - K_P x_p + w$ |
| Control and output constraints (2) | $\begin{array}{l} u^{min} \leq u \leq u^{max} \\ z_{lim}^{min} \leq z_{lim} \leq z_{lim}^{max} \end{array} \Leftrightarrow h(x) = \begin{bmatrix} h_{min}(x) \\ h_{max}(x) \end{bmatrix} = \begin{bmatrix} y_{lim}^{min} - C_{lim} x \\ C_{lim} x - y_{lim}^{max} \end{bmatrix} \leq 0$ <br> $C_{lim} = \begin{bmatrix} -K_I & -K_P \\ 0_{m \times m} & C_{p\,lim} \end{bmatrix}, \quad y^{max} = \begin{bmatrix} u^{max} \\ z_{lim}^{max} \end{bmatrix}, \quad y^{min} = \begin{bmatrix} u^{min} \\ z_{lim}^{min} \end{bmatrix}$ |
| Modified control and output constraints (14) | $\begin{bmatrix} H_{min}(x, u_{bl}, v, w) \\ H_{max}(x, u_{bl}, v, w) \end{bmatrix} = \begin{bmatrix} -H_u \\ H_u \end{bmatrix} \begin{bmatrix} v \\ w \end{bmatrix} + \begin{bmatrix} \Delta H_{min}(x, u_{bl}) \\ \Delta H_{max}(x, u_{bl}) \end{bmatrix} \leq 0$ <br> $\Delta H_{min}(x, u_{bl}) = -H_x x - H_u \begin{bmatrix} -y_{cmd} \\ u_{bl} \end{bmatrix} + \alpha_\pi y_{lim}^{min}$ <br> $\Delta H_{max}(x, u_{bl}) = H_x x + H_u \begin{bmatrix} -y_{cmd} \\ u_{bl} \end{bmatrix} - \alpha_\pi y_{lim}^{max}$ |
| Auxiliary matrices (17), (18) | $H_u = \begin{bmatrix} -K_I & -K_I D_{p\,reg} - K_P B_p \\ 0_{m \times m} & H_w \end{bmatrix},$ <br> $H_x = \begin{bmatrix} (C_{lim})_1 \prod_{j=1}^{r_1} (A - \lambda_{1j} I_n) \\ \vdots \\ (C_{lim})_{2m} \prod_{j=1}^{r_{2m}} (A - \lambda_{2m\,j} I_n) \end{bmatrix}, \quad H_w = \begin{bmatrix} (C_{p\,lim})_1 A_p^{r_{m+1}-1} \\ \vdots \\ (C_{p\,lim})_m A_p^{r_{2m}-1} \end{bmatrix} B_p$ <br> $\alpha_\pi = \mathrm{diag}\left( \begin{bmatrix} c_{10} & c_{20} & \cdots & c_{2m0} \end{bmatrix} \right), \quad c_{i0} = \prod_{j=1}^{r_i} (-\lambda_{ij}) > 0 \quad \forall i = 1, \ldots, 2m$ |
| Min-norm optimal control augmentation (21) | $\begin{bmatrix} v \\ w \end{bmatrix} = H_u^{-1} \left( \max\left( 0_{2m \times 1}, \Delta H_{min}(x, u_{bl}) \right) - \max\left( 0_{2m \times 1}, \Delta H_{max}(x, u_{bl}) \right) \right)$ |

**Table I. Min-norm Optimal CBF-based Servo-Control Augmentation Design Summary.**

A few remarks are in order.

***Remark 3***. The control design method in this paper avoids heuristics-based constraint modifications such as hard saturation function often used to bound a control input to its limits,

$$\mathrm{sat}_{u_{min}}^{u_{max}}(u(t)) = \max\left( u_{min}, \min\left( u(t), u_{max} \right) \right). \tag{22}$$

Using such a saturation heuristic, the dynamical system becomes open loop during control saturation events [3], [16]. Instead, using the control formulation in this paper, the closed-loop system remains closed loop even when control limits have been reached.

*Remark* **4**. An advantage of this formulation is the handling of infeasibilities if they occur. Infeasibilities are often unavoidable during the operation of a dynamical system in a real environment due to unmodeled dynamics, noise, disturbances, etc. Optimization-based solutions often rely on heuristics to handle such infeasibilities to obtain an executable control action [17]. As we only use the optimization-based for controller synthesis that results in an analytical solution, such unavoidable infeasibilities are handled systematically.

*Remark* **5**. The polynomials in (12) control the tradeoff between CBF design conservatism and closed-loop tracking performance, which can be tuned based on performance or robustness objective, e.g. using automated tools as in [18].

*Remark* **6**. The min-norm optimal signal $v(t)$ in (50) represents an AW modification signal, which changes the achievable output tracking command $y_{\text{cmd}}$ into $(y_{\text{cmd}} + v(t))$ and drives the control input $u(t)$ to the prescribed minimum or the maximum total bound, $u^{\min}$ or $u^{\max}$.

*Remark* **7**. Without limited output constraints, only the augmentation signal $v(t)$ is needed for control design, which provides systematic AW protection with

$$v(t) = K_I^{-1}\left(\max\left(0_{m\times 1}, \Delta G_{\min}(x(t))\right) - \max\left(0_{m\times 1}, \Delta G_{\max}(x(t))\right)\right)$$
$$\Delta G_{\min}(x(t)) = \begin{bmatrix} K_I \alpha_\pi & K_I \alpha_\pi D_{p\,\text{reg}} + \left(-K_I C_{p\,\text{reg}} - K_P(A_p + \alpha_\pi)\right)B_p \end{bmatrix} x(t) + K_I y_{\text{cmd}} + \alpha_\pi y_{\lim}^{\min} \quad (23)$$
$$\Delta G_{\max}(x(t)) = -\begin{bmatrix} K_I \alpha_\pi & K_I \alpha_\pi D_{p\,\text{reg}} + \left(-K_I C_{p\,\text{reg}} - K_P(A_p + \alpha_\pi)\right)B_p \end{bmatrix} x(t) - K_I y_{\text{cmd}} + \alpha_\pi y_{\lim}^{\max}.$$

## IV. Analyzability and Robustness of Closed-Loop System

This section derives MIMO margins of the closed-loop system at the input breakpoint based on a switching signal that is driven by active constraints. First, we reformulate the control augmentation solution in (21) as follow. Let

$$\delta(x(t)) = \begin{bmatrix} \delta_1(x(t)) & \cdots & 0 \\ \vdots & \ddots & \vdots \\ 0 & \cdots & \delta_{2m}(x(t)) \end{bmatrix} \in R^{2m \times 2m} \quad (24)$$

defining a diagonal positive-definite state-dependent matrix, whose state-dependent non-negative binary-valued diagonal elements are given by the conditions in (21)

$$\delta_i(x(t)) = \begin{cases} 1, & \text{if } \left[\left(-H_x x(t) - H_u \begin{bmatrix} -y_{\text{cmd}} \\ u_{\text{bl}}(t) \end{bmatrix} + \alpha_\pi y_{\lim}^{\min}\right)_i > 0\right] \vee \left[\left(H_x x(t) + H_u \begin{bmatrix} -y_{\text{cmd}} \\ u_{\text{bl}}(t) \end{bmatrix} - \alpha_\pi y_{\lim}^{\max}\right)_i > 0\right] \\ 0, & \text{otherwise} \end{cases} \quad (25)$$

$\forall i = 1, \ldots, 2m$. Hence, the notation $\delta_i(x(t))$ represents a continuous switching logic based on active constraints, and, by definition, $\|\delta(x(t))\| \leq 1$ uniformly in $x(t)$. Suppose that the baseline controller is designed in the state-proportional feedback form,

$$u_{\text{bl}}(t) = -K_x x(t) \quad (26)$$

with the feedback gain $K_x = \begin{bmatrix} K_I & K_p \end{bmatrix} \in R^{m \times n}$ selected to make the corresponding closed-loop system matrix Hurwitz. Then,

$$\begin{bmatrix} v(t) - y_{\text{cmd}} \\ u_{\text{bl}}(t) + w(t) \end{bmatrix} = \begin{bmatrix} -y_{\text{cmd}} \\ u_{\text{bl}}(t) \end{bmatrix} - H_u^{-1} \delta(x(t)) \left( H_x x(t) + H_u \begin{bmatrix} -y_{\text{cmd}} \\ u_{\text{bl}}(t) \end{bmatrix} - \alpha_\pi y_{\lim}^{\min/\max} \right)$$
$$= \underbrace{\left(I_{(2m)} - H_u^{-1} \delta(x(t)) H_u\right) \begin{bmatrix} -y_{\text{cmd}} \\ u_{\text{bl}}(t) \end{bmatrix}}_{\text{Modified Baseline Control}} - \underbrace{H_u^{-1} \delta(x(t)) H_x x(t)}_{\text{CBF Feedback}} + \underbrace{H_u^{-1} \delta(x(t)) \alpha_\pi y_{\lim}^{\min/\max}}_{\text{CBF Command}} \quad (27)$$

where the $i^{th}$ component of the constant CBF command vector $y_{\lim}^{\min/\max} \in R^{2m}$ is defined as

$$\left(y_{\lim}^{\min/\max}\right)_i = \begin{cases} \left(y_{\lim}^{\min}\right)_i, & \text{if } \left(-H_x x(t) - H_u \begin{bmatrix} -y_{\text{cmd}} \\ u_{\text{bl}}(t) \end{bmatrix} + \alpha_\pi y_{\lim}^{\min}\right)_i > 0 \\ \left(y_{\lim}^{\max}\right)_i, & \text{if } \left(H_x x(t) + H_u \begin{bmatrix} -y_{\text{cmd}} \\ u_{\text{bl}}(t) \end{bmatrix} - \alpha_\pi y_{\lim}^{\max}\right)_i > 0 \\ 0, & \text{otherwise} \end{cases} \tag{28}$$

Within the servo-control design framework, gain and phase margins need to be analyzed at the total control input breakpoint, where the baseline control $u_{\text{bl}}(t)$ and the CBF augmentation signal $w(t)$ are added to form the total control input $u(t)$ to the system. Using the notation in (27), the closed-loop system (6) can be written in terms of the continuous switching logic

$$\dot{x}(t) = Ax + B\left(\left(I_{(2m)} - H_u^{-1}\delta(x(t))H_u\right)\begin{bmatrix} -y_{\text{cmd}} \\ u_{\text{bl}}(t) \end{bmatrix} - H_u^{-1}\delta(x(t))H_x x(t) + H_u^{-1}\delta(x(t))\alpha_\pi y_{\lim}^{\min/\max}\right)$$
$$= Ax(t) + B\begin{bmatrix} -y_{\text{cmd}} \\ -K_x x(t) \end{bmatrix} + BH_u^{-1}\delta(x(t))\left(H_u\begin{bmatrix} y_{\text{cmd}} \\ K_x x(t) \end{bmatrix} - H_x x(t) + \alpha_\pi y_{\lim}^{\min/\max}\right), \tag{29}$$

whereas the total control input becomes

$$u(t) = \begin{bmatrix} 0_{m\times m} & I_m \end{bmatrix}\left(\left(I_{(2m)} - H_u^{-1}\delta(x(t))H_u\right)\begin{bmatrix} -y_{\text{cmd}} \\ -K_x x(t) \end{bmatrix} - H_u^{-1}\delta(x)H_x x(t) + H_u^{-1}\delta(x(t))\alpha_\pi y_{\lim}^{\min/\max}\right)$$
$$= \begin{bmatrix} 0_{m\times m} & I_m \end{bmatrix}\left(\begin{bmatrix} -y_{\text{cmd}} \\ -K_x x(t) \end{bmatrix} - H_u^{-1}\delta(x(t))H_u\begin{bmatrix} -y_{\text{cmd}} \\ -K_x x(t) \end{bmatrix} - H_u^{-1}\delta(x)H_x x(t) + H_u^{-1}\delta(x(t))\alpha_\pi y_{\lim}^{\min/\max}\right) \tag{30}$$
$$= -K_x x(t) + \delta_w(x(t))\left(H_u\begin{bmatrix} y_{\text{cmd}} \\ K_x x(t) \end{bmatrix} - H_x x(t) + \alpha_\pi y_{\lim}^{\min/\max}\right)$$

with

$$\delta_w(x(t)) = \begin{bmatrix} 0_{m\times m} & I_m \end{bmatrix} H_u^{-1}\delta(x(t)). \tag{31}$$

The system loop gain transfer function matrix $L_u(s)$ can be computed with the total servo-controller in the form of (27), while zeroing out command terms $y_{\text{cmd}}$ and $y_{\lim}^{\min/\max}$,

$$u_{\text{out}}(t) = -K_x x(t) + \delta_w(x(t))\left(H_u\begin{bmatrix} 0_{m\times m} \\ I_m \end{bmatrix}K_x - H_x\right)x(t) = -\left(K_x - \delta_w(x(t))\left(H_u\begin{bmatrix} 0_{m\times m} \\ I_m \end{bmatrix}K_x - H_x\right)\right)x(t). \tag{32}$$

For this analysis, all active AW augmentation loops due to $v(t)$ need to be closed, since the integrator dynamics represent a known part of the total controller. Hence, we need to separate the AW signal $v(t)$ from the control augmentation signal $w(t)$ for robustness analysis, where we define

$$v(t) = -\delta_v(x(t))\left(H_u\begin{bmatrix} 0 \\ I \end{bmatrix}K_x + H_x\right)x(t) \tag{33}$$

with

$$\delta_v(x(t)) = [I_m \quad 0_{m \times m}] H_u^{-1} \delta(x(t)). \tag{34}$$

Thus, we write the closed-loop system dynamics as

$$\dot{x}(t) = A x(t) + B \begin{bmatrix} v(t) \\ u(t) \end{bmatrix}$$

$$= \left( A - \underbrace{B \begin{bmatrix} I_m \\ 0_{m \times m} \end{bmatrix} \delta_v(x(t)) \left( H_u \begin{bmatrix} 0_{m \times m} \\ I_m \end{bmatrix} K_x + H_x \right)}_{A_v(\delta_v(x(t)))} \right) x(t) + B \begin{bmatrix} 0_{m \times m} \\ I_m \end{bmatrix} u(t) \tag{35}$$

with

$$A_v(\delta_v(x(t))) = B \begin{bmatrix} I_m \\ 0_{m \times m} \end{bmatrix} \delta_v(x(t)) \left( H_u \begin{bmatrix} 0_{m \times m} \\ I_m \end{bmatrix} K_x + H_x \right). \tag{36}$$

Finally, we can state the loop gain

$$u_{\text{out}}(t) = -\underbrace{\left( K_x - \delta_w \left( H_u \begin{bmatrix} 0_{m \times m} \\ I_m \end{bmatrix} K_x - H_x \right) \right) (s I_n - (A - A_v(\delta_v)))^{-1} B \begin{bmatrix} 0_{m \times m} \\ I_m \end{bmatrix}}_{\text{Loop Gain:} L_u(s;\delta)} u_{\text{in}}(t) = -L_u(s;\delta) u_{\text{in}}(t). \tag{37}$$

In this case, SISO and MIMO margins at the system input breakpoint are defined based on the resulting $(m \times m)$–dimensional loop gain transfer function matrix, parameterized with the binary-valued matrix $\delta$,

$$L_u(s;\delta) = \left( K_x - \delta_w \left( H_u \begin{bmatrix} 0_{m \times m} \\ I_m \end{bmatrix} K_x - H_x \right) \right) (s I_n - (A - A_v(\delta_v)))^{-1} B \begin{bmatrix} 0_{m \times m} \\ I_m \end{bmatrix} \tag{38}$$

where $\delta_v$ and $\delta_w$ can be simplified using (18),

$$\delta_v = [I_m \quad 0_{m \times m}] H_u^{-1} \delta = [I_m \quad 0_{m \times m}] \begin{bmatrix} -K_I^{-1} & -K_I^{-1} K_x B H_w^{-1} \\ 0_{m \times m} & H_w^{-1} \end{bmatrix} \delta = \begin{bmatrix} -K_I^{-1} & -K_I^{-1} K_x B H_w^{-1} \end{bmatrix} \delta$$

$$\delta_w = [0_{m \times m} \quad I_m] H_u^{-1} \delta = [0_{m \times m} \quad I_m] \begin{bmatrix} -K_I^{-1} & -K_I^{-1} K_x B H_u^{-1} \\ 0_{m \times m} & H_w^{-1} \end{bmatrix} \delta = \begin{bmatrix} 0_{m \times m} & H_w^{-1} \end{bmatrix} \delta. \tag{39}$$

Hence, the loop gain (38) parameterized with $\delta$ can be used to compute SISO and MIMO gain and phase margins [1] for all possible combinations of the binary-valued diagonal elements of $\delta$. For relative stability analysis, it is assumed that $\delta$ is a constant diagonal matrix, with binary values on its diagonal. Also, only the feedback portion of the CBF controller needs to be considered and in that case, relative stability metrics, such as gain and phase margins, can be computed and analyzed based on the overall system block-diagram, shown in Fig. 1.

**Remark 8**. It is interesting to note that the CBF augmentation (21) can be viewed as a continuous piece-wise linear state feedback linearizing controller [15], [19], with the embedded switching logic. In other words, (21) can be viewed as a dynamic inversion (DI) controller, but only once constraints become active. The DI nature is due to the feedback gain $(H_u^{-1} \delta(x) H_x)$ and because of that, the corresponding closed-loop system can also be analyzed within the DI framework.

## V. Control Augmentation Insights

### A. Limited Output Constraint with Proportional Controller
Consider a scalar system,

$$\dot{x}_p(t) = a_p x_p(t) + b_p u(t) \tag{40}$$

with a baseline proportional feedback controller,

$$u_{bl}(t) = -k x_p(t) \tag{41}$$

and suppose that the state feedback gain $k$ is chosen to stabilize the system dynamics $a_{p\,cl} = a_p - b_p k < 0$. The total control command is

$$u(t) = u_{bl}(t) + w(t) = -k x_p(t) + w(t) \tag{42}$$

with the control augmentation signal $w(t)$ designed to enforce the box operational limits (2) on the system state,

$$w(t) = H_w^{-1}\left(\max\left(0, \Delta H_{\min}(x_p(t))\right) - \max\left(0, \Delta H_{\max}(x_p(t))\right)\right). \tag{43}$$

In this case, the limited output is the system state,

$$x_p^{\min}(t) \leq z_{\lim}(t) = x_p(t) \leq x_p^{\max} \quad \Leftrightarrow \quad h(x_p(t)) = \begin{bmatrix} h_{\min}(x_p(t)) \\ h_{\max}(x_p(t)) \end{bmatrix} = \begin{bmatrix} z_{\lim}^{\min} - z_{\lim}(t) \\ z_{\lim}(t) - z_{\lim}^{\max} \end{bmatrix} = \begin{bmatrix} x_p^{\min} - x_p(t) \\ x_p(t) - x_p^{\max} \end{bmatrix} \leq 0 \tag{44}$$

and the corresponding input-to-output relative degree is one. Then, $H_w = b_p$,

$$\Delta H_{\min}(x_p(t)) = \left(\dot{h}_{\min}(x_p(t))\right)_{u=u_{bl}} + c_0 h_{\min}(x_p(t)) = -a_{p\,cl} x_p(t) + c_0\left(x_p^{\min} - x_p(t)\right)$$

$$\Delta H_{\max}(x_p(t)) = \left(\dot{h}_{\max}(x_p(t))\right)_{u=u_{bl}} + c_0 h_{\max}(x_p(t)) = a_{p\,cl} x_p(t) + c_0\left(x_p(t) - x_p^{\max}\right). \tag{45}$$

In (45), $c_0 = (-\lambda_1) > 0$ is the negative of the stable real eigenvalue $\lambda_1 < 0$ of the first order polynomial (12). Clearly, the two output constraints (45) are mutually exclusive and only one can take place at any given time. In other words, only one of the signals can become positive, but not both.

Suppose that the max output constraint becomes active. Then,

$$w(t) = -\frac{1}{b_p} \Delta H_{\max}(x_p(t)) = -\frac{1}{b_p}\left(a_{p\,cl} x_p(t) + c_0\left(x_p(t) - x_p^{\max}\right)\right) \tag{46}$$

and the closed-loop system dynamics with active limited output constraint become

$$\dot{x}_p(t) = a_{p\,cl} x_p(t) + b_p w(t) = a_{p\,cl} x_p(t) + b_p\left(-\frac{1}{b_p}\left(a_{p\,cl} x_p(t) + c_0\left(x_p(t) - x_p^{\max}\right)\right)\right)$$

$$= a_{p\,cl} x_p(t) - \left(a_{p\,cl} x_p(t) + c_0\left(x_p(t) - x_p^{\max}\right)\right) = -c_0\left(x_p(t) - x_p^{\max}\right). \tag{47}$$

Hence, it is easy to see that the control augmentation forces the closed-loop system dynamics to evolve with the differential equation defined by the boundary of the active constraint. Closed-loop stability is guaranteed by construction with eigenvalues $-c_0 = \lambda_1 < 0$ and the operational limit $x_p^{\max}$ is being approached asymptotically.

Overall, the closed-loop system dynamics for the scalar system are

$$\dot{x}_p(t) = \begin{cases} -c_0\left(x_p(t) - x_p^{\min}\right) & \text{if } \Delta H_{\min}(x_p(t)) \geq 0 \\ -c_0\left(x_p(t) - x_p^{\max}\right) & \text{if } \Delta H_{\max}(x_p(t)) \geq 0 \\ a_{p\,cl} x_p(t) & \text{else.} \end{cases} \tag{48}$$

### B. Input Constraint and Systematic Anti-Windup with Proportional-Integral (PI) Controller

Consider a SISO system,

$$\dot{x}_p(t) = A_p x_p(t) + b_p u(t) \tag{49}$$

and a baseline proportional-integral feedback controller with control augmentation $v(t)$ to be designed such that control input constraints are satisfied for all times and to keep the integrated output tracking error ultimately bounded,

$$u(t) = -k_p x_p(t) - k_I e_{yI}(t)$$
$$\dot{e}_{yI}(t) = c_{p\,reg} x_p(t) - y_{cmd} + v(t) \tag{50}$$

and

$$u^{min} \leq u(t) \leq u^{max} \Leftrightarrow \begin{bmatrix} u(t) - u^{max} \\ u^{min} - u(t) \end{bmatrix} \leq 0. \tag{51}$$

Let $\alpha_u > 0$ be used for the derivation of its modified constraint for controller design, i.e.,

$$\dot{u}(t) + \alpha_u \left( u(t) - u^{max} \right) = -k_p \left( A_p x_p(t) + b_p u(t) \right) - k_I \left( c_{p\,reg} x_p(t) - y_{cmd} + v(t) \right) + \alpha_u \left( u(t) - u^{max} \right) \leq 0$$
$$-\dot{u}(t) + \alpha_u \left( u^{min} - u(t) \right) = +k_p \left( A_p x_p(t) + b_p u(t) \right) + k_I \left( c_{p\,reg} x_p(t) - y_{cmd} + v(t) \right) + \alpha_u \left( u^{min} - u(t) \right) \leq 0. \tag{52}$$

Expression (52) represents a function that is linear in control decision variables with $\alpha_u > 0$. Suppose the min control input constraint becomes active. Then, the control augmentation produces values that strictly enforce the control constraint's boundary

$$\dot{u}(t) + \alpha_u \left( u(t) - u^{min} \right) = 0. \tag{53}$$

Hence, (53) shows that, when the min control input constraint becomes active, the control input follows a differential equation, and the closed-loop system with active control input constraint becomes

$$\begin{bmatrix} \dot{x}_p(t) \\ \dot{u}(t) \end{bmatrix} = \begin{bmatrix} A_p & b_p \\ 0 & -\alpha_u \end{bmatrix} \begin{bmatrix} x_p(t) \\ u(t) \end{bmatrix} + \begin{bmatrix} 0 \\ 1 \end{bmatrix} u^{min} \tag{54}$$

with stable eigenvalues $(-\alpha_u) < 0$ and $\lambda_i(A_p) < 0$ due to $A_p$ being Hurwitz. In order to show boundedness for the integrated output tracking error, consider the coordinate transformation

$$\begin{bmatrix} x_p(t) \\ u(t) \end{bmatrix} = \begin{bmatrix} I_{n_p} & 0 \\ -k_p & -k_I \end{bmatrix} \begin{bmatrix} x_p(t) \\ e_{yI}(t) \end{bmatrix} \tag{55}$$

applied to (54) resulting in

$$\begin{bmatrix} \dot{x}_p(t) \\ \dot{e}_{yI}(t) \end{bmatrix} = A_G \begin{bmatrix} x_p(t) \\ e_{yI}(t) \end{bmatrix} + b_G u^{min}, \quad A_G = \begin{bmatrix} A_p - b_p k_p & -b_p k_I \\ k_p k_I^{-1} \left( A_p - b_p k_p - \alpha_v I_{n_p} \right) & -k_p b_p - \alpha_u \end{bmatrix}, b_G = \begin{bmatrix} 0 \\ -k_I^{-1} \end{bmatrix}. \tag{56}$$

Since the coordinate transformation in (55) does not change the eigenvalues of the closed-loop system, $A_G$ is Hurwitz with the stable eigenvalues $(-\alpha_u) < 0$ and $\lambda_i(A_p) < 0$. This implies that the error state remains UUB once constraints become active.

Overall, in the absence of limited output constraints, the closed-loop system dynamics are given by

$$\begin{bmatrix} \dot{x}_p(t) \\ \dot{e}_{yI}(t) \end{bmatrix} = \begin{cases} A_G \begin{bmatrix} x_p(t) \\ e_{yI}(t) \end{bmatrix} + b_G u^{min} & \text{if modified minimum input constraint is active} \\ A_G \begin{bmatrix} x_p(t) \\ e_{yI}(t) \end{bmatrix} + b_G u^{max} & \text{if modified maximum input constraint is active} \\ \begin{bmatrix} A_p & b_p \\ c_{p\,reg} & 0 \end{bmatrix} \begin{bmatrix} x_p(t) \\ e_{yI}(t) \end{bmatrix} + \begin{bmatrix} 0 \\ -1 \end{bmatrix} y_{cmd} & \text{else.} \end{cases} \tag{57}$$

Fig. 2 illustrates the control input constraint and the notion of modified constraints. In [8], a similar term but with a positive constant $\alpha_u$ is introduced and called the "negative margin," indicating that its purpose is to repel the system trajectories away from their designated limit boundaries. Fig. 2 shows a baseline controller without modifications, with hard saturations, and with the control augmentation in this paper. The control augmentation is triggered and intervenes once the baseline control input is getting close to the control limits but also considers how fast the control limit is being approached. This tradeoff between distance and gradient is controlled by the "slope" parameter $\alpha_u$ in the modified inequality formulation in (14).

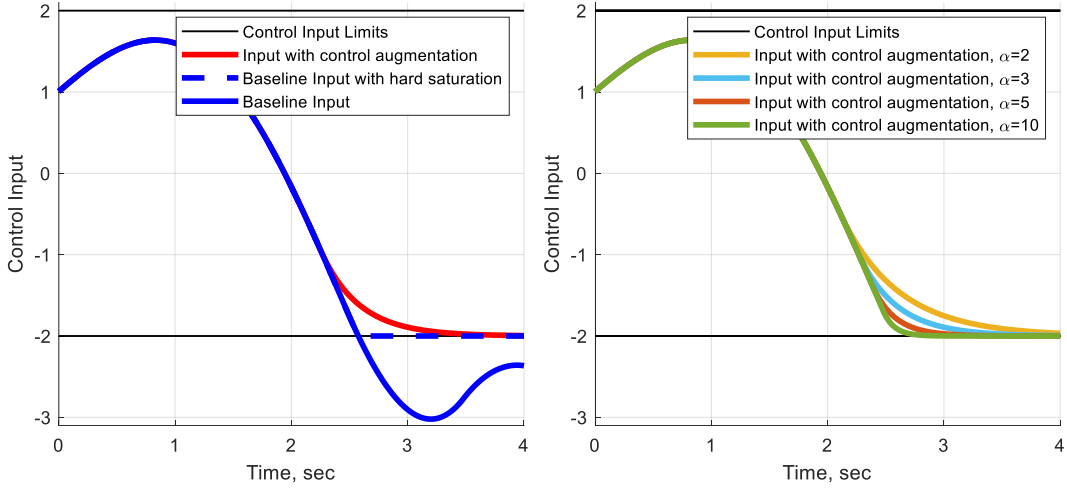

**Fig. 2. CBF Augmentation illustration for control input constraint. Left: Comparison of baseline controller without modifications (blue), with hard saturation (blue dashed), and with control augmentation (red) computed by the proposed approach. Right: Control input commands for different design parameters. The control augmentation logic intervenes if the baseline controller is approaching the boundary and would violate the constraint. Once the control augmentation is triggered, the control input follows the differential equation defined by the boundary of the active constraint, i.e., the control input converges asymptotically to the limit.**

## VI. Flight Control Trade Studies

### A. Linear Aircraft Model

First, simulation results are presented using a linear aircraft model. Consider the roll-yaw dynamics representative of a mid-size aircraft, (see [1], Section 14.8, pp. 622–626),

$$\dot{x}_p(t) = A_p x_p(t) + B_p u(t), \tag{58}$$

where the state $x_p = \begin{bmatrix} \beta & p_s & r_s \end{bmatrix}^T$ consists of the sideslip angle $\beta$ (rad), as well as the vehicle stability axis roll and yaw rates (rad/sec), $p_s$ and $r_s$. The control input $u = \begin{bmatrix} \delta_{ail} & \delta_{rud} \end{bmatrix}^T$ is represented by the aileron and the rudder deflections (rad), $\delta_{ail}$ and $\delta_{rud}$. The regulated output is the roll rate $p_s$ (rad/sec) and the lateral load factor $N_y$ (g-s), where $g = 32.174$ is the gravitational acceleration (ft/sec²),

$$y_{reg}(t) = \begin{bmatrix} p_s(t) \\ N_y(t) \end{bmatrix} = C_{p\,reg} x_p(t) + D_{p\,reg} u(t). \tag{59}$$

The aircraft model data are computed using numerical linearization with respect to al 1g-level flight trim (i.e., equilibrium) at the flight conditions, $V_0 = 717.17 \text{ft/sec}$, $25.000 \text{ft}$ altitude, and $4.5627 \text{deg}$ angle of attack, which provides the following linearized system dynamics model:

$$A_p = \begin{bmatrix} -0.11794 & 0.00085 & -1.0001 \\ -7.0113 & -1.4492 & 0.22059 \\ 6.3035 & 0.06511 & -0.41172 \end{bmatrix}, B_p = \begin{bmatrix} 0 & 0.015257 \\ -7.9662 & 2.6875 \\ 0.60926 & -2.3577 \end{bmatrix} \quad (60)$$

$$C_{p\,\text{reg}} = \begin{bmatrix} 0 & 1 & 0 \\ -2.6049 & 0.018724 & 0.067695 \end{bmatrix}, D_{p\,\text{reg}} = \begin{bmatrix} 0 & 0 \\ 0 & 0.33698 \end{bmatrix}.$$

A baseline Linear Quadratic Regulator (LQR) PI controller [1] is designed without operational limits, using the integrated output tracking error dynamics,

$$\dot{e}_{yI}(t) = y_{\text{reg}}(t) - y_{\text{cmd}} = \begin{bmatrix} p_s(t) - p_{s\,\text{cmd}} \\ N_y(t) - N_{y\,\text{cmd}} \end{bmatrix} \quad (61)$$

and the following LQR weights

$$Q_{\text{lqr}} = \text{diag}(1.025 \quad 1.0289 \quad 0 \quad 0 \quad 1.6021), R_{\text{lqr}} = \text{diag}(1 \quad 0.49129). \quad (62)$$

The operational constraints are given by the box constraints illustrated in Fig. 3, i.e., the aileron and the rudder position limits are set to ±3deg and ±2deg, respectively, while the roll rate and the sideslip angle limits are ±18deg/s and ±0.5deg, respectively. The regulated outputs are a zero lateral load factor and roll rate commands that switch between 15deg/s, 0deg/s, -15deg/s, and 0deg/s every ten seconds. In order to achieve a coordinated turn, the lateral acceleration command switches between 0.0312g, 0g, -0.0312g, 0g in the same frequency. The selection of these limits allows to demonstrate efficiency of the norm-based control augmentation with respect to control input and limited output constraints. The "slopes" of the constraints corresponding to aileron, rudder, roll rate, and sideslip angle are designed using the following four values in the diagonal matrix,

$$\alpha_\pi = \text{diag}(80 \quad 8 \quad 40 \quad 40). \quad (63)$$

*1. Baseline Controller without Saturation*

For comparison, Fig. 3 shows a baseline controller without control augmentation and without a saturation. The baseline controller does not keep the rudder and aileron commands within their limits, which is expected due to the absence of any modification. In order to track the commanded roll rate, the baseline controller produces a sideslip angle that exceeds the operational boundaries. The states and tracking error integrators remain bounded, which are displayed on the right in Fig. 3. Here, it can be seen how the roll rate is being tracked with the sideslip angle exceeding its 0.5deg limits. For these reasons, a saturation function is often used in practice to prevent a command from exceeding its operational limits. However, using a saturation function without a systematic treatment of its implications can lead to windup and degradation of performance, which is shown next.

*2. Baseline Controller with Saturation*

Fig. 4 shows the same baseline controller, but with a hard saturation on the aileron and rudder command as implemented in (22). This leads to input constraint satisfaction, however, due to the lack of systematic treatment of such control saturation, the sideslip angle output constraint is violated, see bottom left plot and top right plot in Fig. 4. Further, the integrator state of the lateral acceleration winds up, see bottom right plot. While the roll rate command is tracked, this windup leads to undesirable performance, which manifests itself best in the yaw rate with heavy oscillations in Fig. 4. Further, the dynamical system becomes open loop during control saturation. In practice, heuristics-based AW modifications are often used [3], however, such modifications can lack a control-theoretic foundation.

*3. Controller with Proposed Augmentation*

Fig. 5 shows the operational limits for the controller with the proposed CBF augmentation. Similar to using a saturation function, all input control bounds are satisfied. However, the proposed control augmentation also prevents output constraints from being violated. Further, the augmentation command prevents integrator windup by construction once the control input boundary is reached, see bottom right plot in Fig. 5. This behavior is caused by adjusting the control augmentation signal $v$ to account for the control input limits systematically in the flight control system. Overall, the control augmentation framework in this paper renders the system trajectories forward invariant with respect to its operational limits.

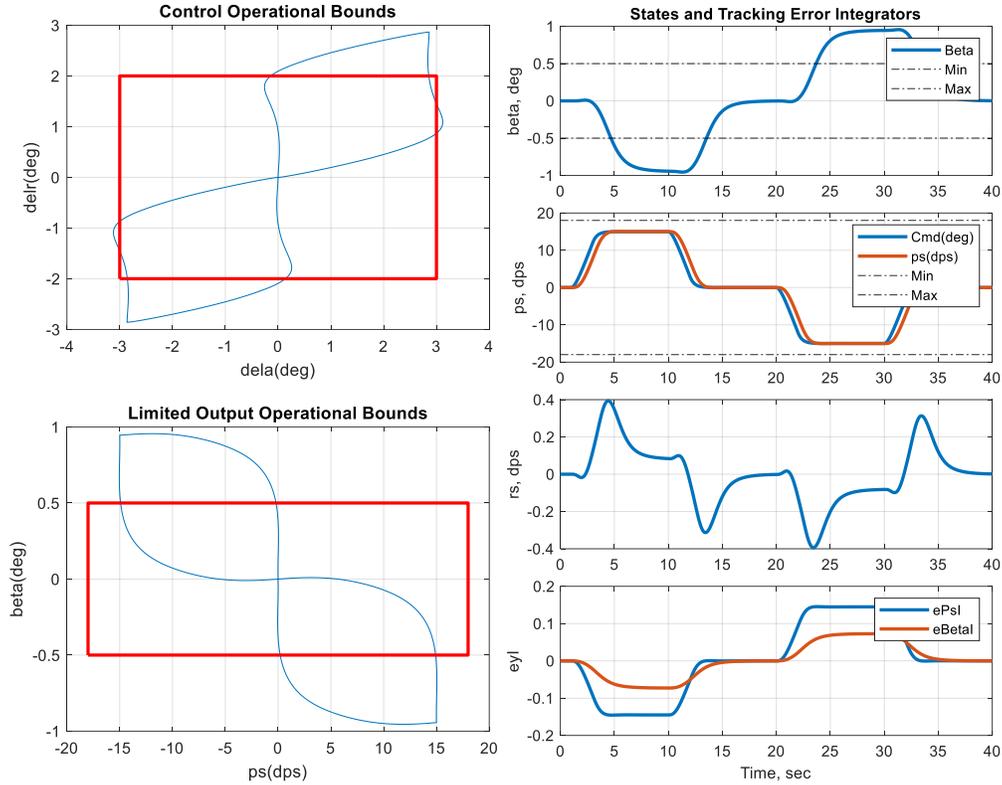

Fig. 3. Coordinated turn with unconstrained baseline LQR PI controller. Left: Operational limits. Right: Aircraft states and integrator states. The baseline controller without saturation exceeds the operational control input limits to achieve the desired roll rate of 15deg/s.

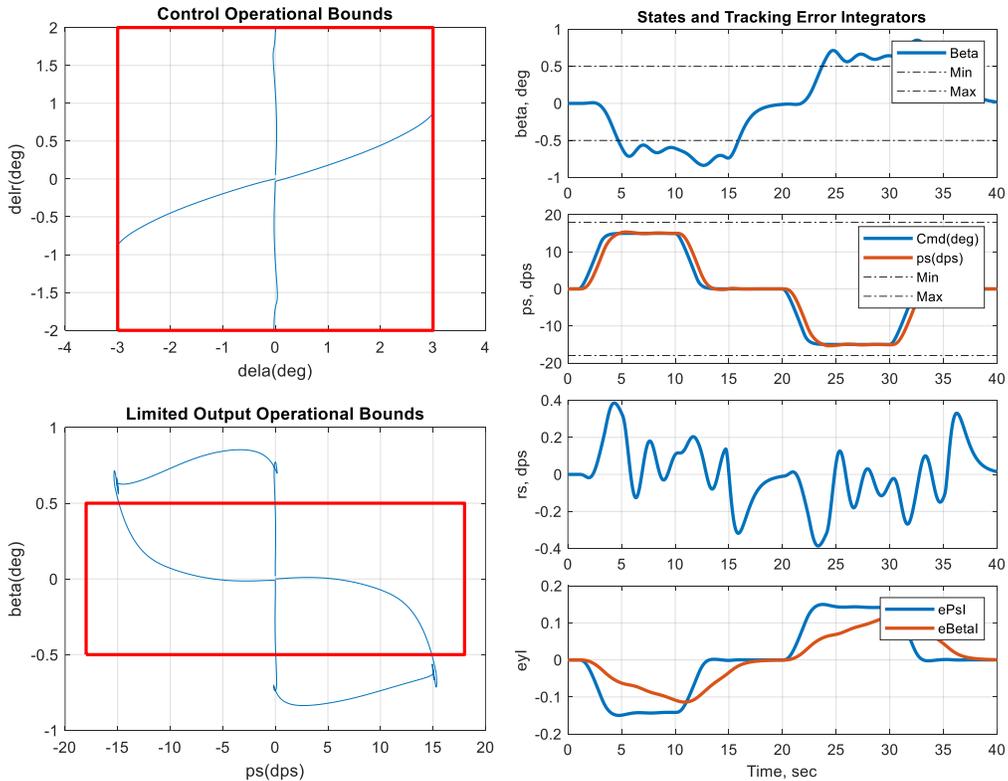

Fig. 4. Coordinated turn with baseline LQR PI controller and hard saturation. Left: Operational limits. Right: Aircraft states and integrator states. A hard saturation on the aileron and rudder control input enforces limits but fails to address output constraints. The bottom right plot shows the windup of the integrator state, which leads to a lag in roll rate responsiveness, see second plot from the top.

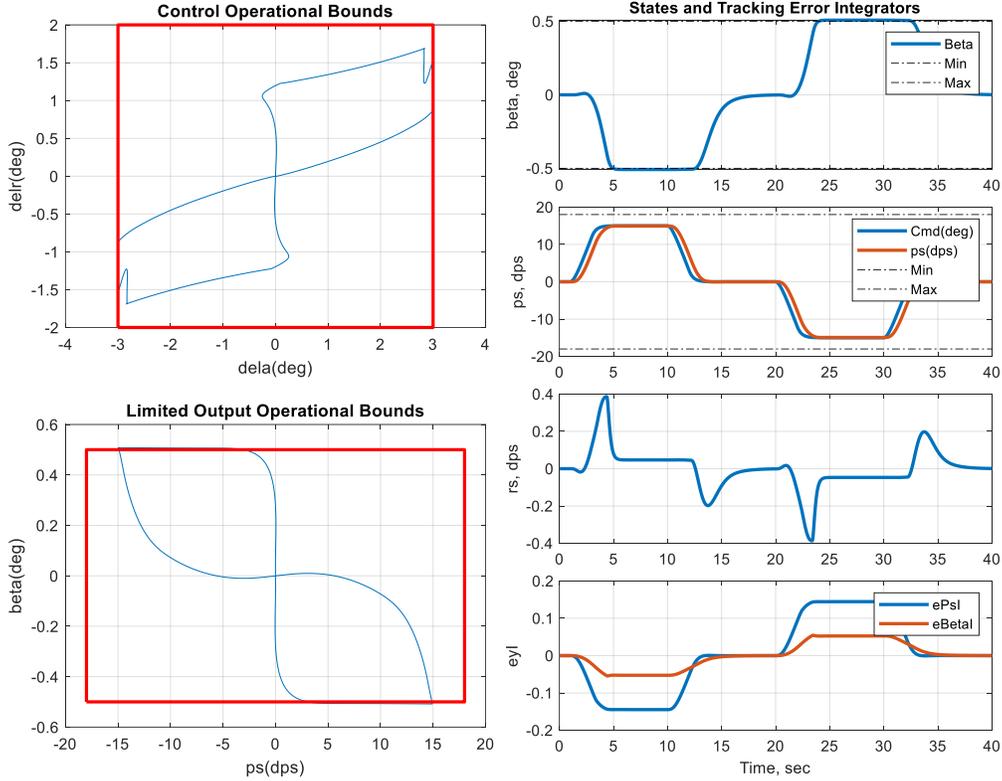

**Fig. 5. Coordinated turn with control augmentation. Left: Operational limits. Right: Aircraft states and integrator states. The control augmentation enforces the aileron limits. Due to the limits on aileron deflections, the commanded roll rate cannot be achieved. However, the control augmentation prevents the windup of the integral error, see bottom right plot.**

### B. Phase and Gain Margin Analysis for Linearized Aircraft Model

One significant advantage of the proposed CBF augmentation design is its analyzability through linear systems theory. To illustrate this, Table II presents the MIMO margins computed at the input breakpoint, specifically focusing on control input constraints. The table compares three scenarios: a baseline controller without augmentation, a baseline controller employing the hard saturation heuristic as described in (22), and the proposed baseline controller with CBF augmentation.

In the absence of active constraints, all three cases yield identical performance, as indicated in the first row of Table II. However, when the aileron control input reaches its limit, the controller utilizing hard saturation becomes unstable as the aileron becomes open loop. In contrast, the proposed method with active aileron constraint remains closed-loop and exhibits comparable margins to the unconstrained case, while ensuring constraint satisfaction, as shown in the second row of Table II. When the rudder control input approaches its limits, the hard saturation controller remains stable but experiences a significant reduction in margins. Conversely, the margins for the proposed method are only slightly affected by the active rudder constraint, see the third row of Table II. Finally, in the scenario where both the aileron and rudder control inputs reach their respective limits simultaneously, the entire plant transitions to an open-loop state if the hard saturation heuristic is employed. In contrast, the proposed method with CBF augmentation maintains a closed-loop configuration, demonstrating only a slight reduction in margins. This analysis underscores the robustness and effectiveness of the proposed CBF augmentation in managing control input constraints while preserving system stability.

|  | Baseline Controller | Baseline Controller with Hard Saturation Heuristic | Baseline Controller with AW Augmentation |
|---|---|---|---|
| No Constraint Active | GM = [-23.4  27.8]dB<br>PM = [-57.3  57.3]deg | GM = [-23.4  27.8]dB<br>PM = [-57.3  57.3]deg | GM = [-23.4  27.8]dB<br>PM = [-57.3  57.3]deg |
| Aileron Constraint Active | N/A (constraint violated) | GM = [0  0]dB<br>PM = [0  0]deg<br>Aileron becomes open loop | GM = [-23.4  28.0]dB<br>PM = [-57.4  57.4]deg |
| Rudder Constraint Active | N/A (constraint violated) | GM = [-5.93  5.82]dB<br>PM = [-28.7  28.7]deg<br>Rudder becomes open loop | GM = [-20.1  27.5]dB<br>PM = [-57.2  57.2]deg |
| Aileron and Rudder Constraints Active | N/A (constraints violated) | GM = [-Inf  Inf]dB<br>PM = [-60  60]deg<br>Plant becomes open loop | GM = [-19.9  27.6]dB<br>PM = [-57.3  57.3]deg |

Table II. Multi-Input Multi-Output (MIMO) margins for baseline controller without constraint enforcement, baseline controller with hard saturation logic, and baseline controller with anti-windup augmentation.

## C. Nonlinear Simulation Study

In this section, we present simulation results derived from a nonlinear six-degree-of-freedom (6DoF) rigid body aircraft model that incorporates actuator dynamics, aerodynamic modeling, atmospheric conditions, and engine performance. Our focus is primarily on the AW component of the control system. The 6DoF model includes an airspeed controller, a pitch controller, and a roll-yaw controller. The pitch and roll-yaw controller utilize a proportional-integral feedback baseline controller augmented by the proposed CBF methodology.

Fig. 6 illustrates the results of roll rate command tracking. Initially, when a roll rate of 10 degrees per second (dps) is commanded, the system operates without reaching any control input limits. As a result, the commanded roll rate is approached rapidly and tracked robustly, see the top left plot from 1 to 5 seconds. During this phase, the AW modification remains inactive, which is reflected in the AW-modified command closely following the commanded roll rate. However, at 9 seconds, a roll rate command of -20 dps is issued. In this scenario, the presence of aileron and rudder actuation limitations leads to a noticeable difference in transient performance compared to the unconstrained case. The AW-modified command signal anticipates the control input limits, resulting in a delay in the roll rate buildup. Despite this delay, the system ultimately approaches and tracks the commanded value of -20 dps, demonstrating the effectiveness of the proposed method in managing external command tracking within the constraints imposed by actuator limitations.

Moreover, while the sideslip angle command remains zero throughout the simulation, the AW-modified sideslip command is nonzero. This adjustment arises from the necessity to compensate for the limitations of the aileron and rudder actuation. To effectively track the roll rate while operating under these constraints, the system augments the command with a small sideslip angle. This strategic modification exemplifies how the proposed CBF augmentation enhances the controller's ability to maintain performance and stability in the presence of operational constraints, ultimately ensuring that the aircraft can respond to external commands as effectively as possible.

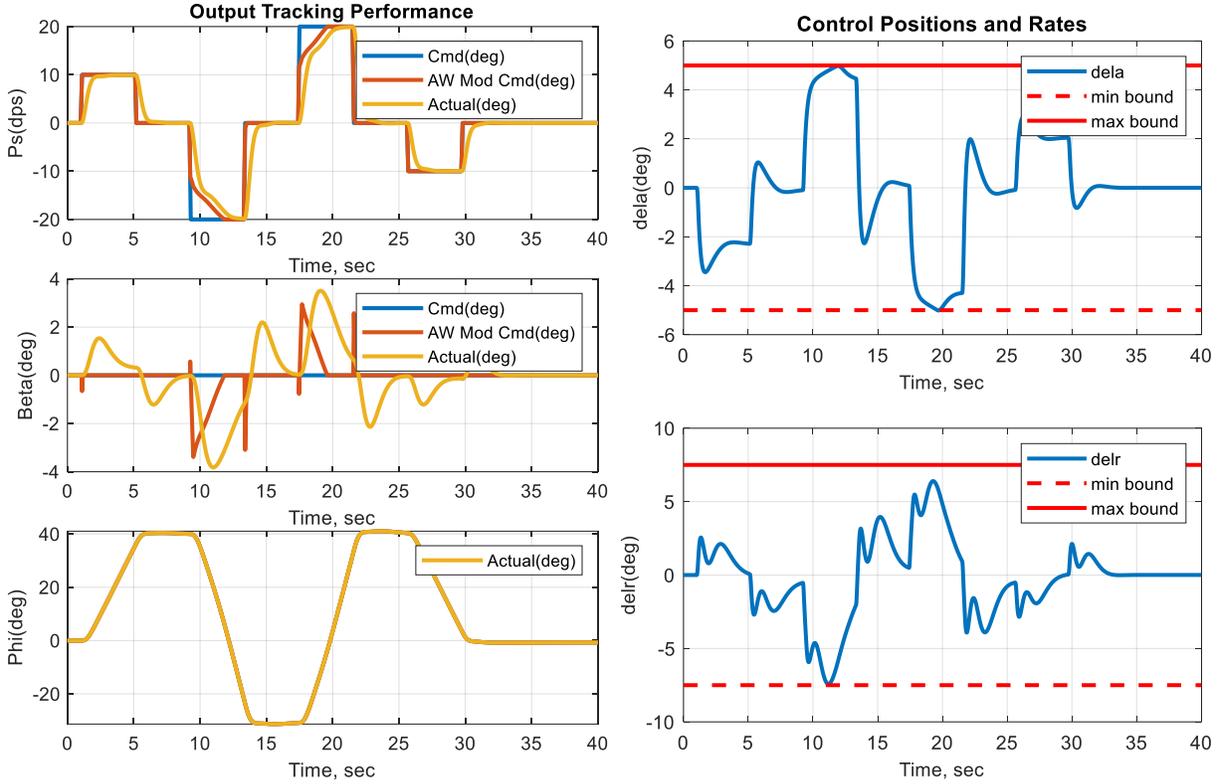

**Fig. 6. Coordinated turn with baseline LQR PI controller and hard saturation. Left: Operational limits. Right: Aircraft states and integrator states.**

## VII.  Conclusions

In this paper, we have presented a systematic approach for designing robust linear servo-controllers that effectively manage both control input and output constraints. Our framework leverages CBFs to ensure safety and performance in dynamic systems, particularly in the context of flight control applications. The simulation results using a linear aircraft model demonstrate accurate tracking of external commands while satisfying operational constraints on both the control input and limited output. Furthermore, the nonlinear simulation results reveal the effectiveness of the anti-windup modification in the presence of unmodeled dynamics, underscoring the applicability of the proposed method to real-world systems.

Additionally, our MIMO margin analysis highlights the closed-loop implementation of the proposed method under constraints. Notably, the gain and phase margins exhibit only minor differences from the baseline controller without active constraints, indicating that the proposed approach maintains robust performance. In contrast, a controller employing hard saturation heuristics becomes open loop with active constraints, resulting in gain and phase margins that fall below the values typically required for safe operation. These findings show the advantages of our systematic control design, providing a reliable framework for enhancing the robustness and safety of FCSs in the presence of operational constraints.

## References

[1] E. Lavretsky and K. A. Wise, Robust adaptive control. In Robust and adaptive control: With aerospace applications, Second Edition, Advanced Textbooks in Control and Signal Processing, Springer Nature Switzerland AG, 2024.

[2] J. Hespanha, Linear systems theory, Princeton university press, 2018.

[3] K. Åström and R. Murray, Feedback systems: an introduction for scientists and engineers, Princeton, NJ: Princeton university press, 2021.

[4] E. Lavretsky and M. Menner, "Servo-Controllers for Linear Time-Invariant Systems with Operational Constraints," in *American Control Conference (ACC)*, 2025.

[5] M. Nagumo, "Über die Lage der Integralkurven gewöhnlicher Differentialgleichungen," *Proceedings of the Physico-Mathematical Society of Japan. 3rd Series,* vol. 24, pp. 551-559, 1942.

[6] M. Menner and E. Lavretsky, "Translation of Nagumo's Foundational Work on Barrier Functions: On the Location of Integral Curves of Ordinary Differential Equations," *arXiv preprint arXiv:2406.18614,* 2024.

[7] A. McNabb, "Comparison theorems for differential equations," *Journal of mathematical analysis and applications,* vol. 119, no. 1-2, pp. 417-428, 1986.

[8] R. Freeman and P. Kokotovic, "Inverse optimality in robust stabilization," *SIAM journal on control and optimization,* vol. 34, no. 4, pp. 1365-1391, 1996.

[9] S. Boyd and L. Vandenberghe, Convex optimization, Cambridge, MA: Cambridge university press, 2004.

[10] A. D. Ames, X. Xu, J. W. Grizzle and P. Tabuada, "Control barrier function based quadratic programs for safety critical systems," *IEEE Trans. Automatic Control,* vol. 62, no. 8, pp. 3861-3876, 2016.

[11] K. Wabersich and M. Zeilinger, "Predictive control barrier functions: Enhanced safety mechanisms for learning-based control," *IEEE Trans. Automatic Control,* vol. 68, no. 5, pp. 2638-2651, 2022.

[12] K. Wabersich, A. Taylor, J. Choi, K. Sreenath, C. Tomlin, A. Ames and M. Zeilinger, "Data-driven safety filters: Hamilton-jacobi reachability, control barrier functions, and predictive methods for uncertain systems," *IEEE Control Systems Magazine,* vol. 43, no. 5, pp. 137-177, 2023.

[13] M. Morari and J. Lee, "Model predictive control: past, present and future," *Computers & chemical engineering,* vol. 23, no. 4-5, pp. 667-682, 1999.

[14] L. Hewing, K. Wabersich, M. Menner and M. Zeilinger, "Learning-based model predictive control: Toward safe learning in control," *Annual Review of Control, Robotics, and Autonomous Systems,* vol. 3, no. 1, pp. 269-296, 2020.

[15] A. Isidori, J. van Schuppen, E. Sontag, M. Thoma and M. Krstic, Communications and control engineering, Berlin, Germany: Springer-Verlag, Berlin, 1995.

[16] S. Tarbouriech and M. Turner, "Anti-windup design: an overview of some recent advances and open problems," *IET control theory & applications,* vol. 3, no. 1, pp. 1-19, 2009.

[17] M. Muehlebach and M. I. Jordan, "On Constraints in First-Order Optimization: A View from Non-Smooth Dynamical Systems," *Journal of Machine Learning Research,* vol. 23, pp. 1-47, 2022.

[18] M. Menner, K. Berntorp and S. Di Cairano, "Automated controller calibration by Kalman filtering," *IEEE Transactions on Control Systems Technology,* vol. 31, no. 6, pp. 2350-2364, 2023.

[19] H. Khalil, Nonlinear Systems (3rd Edition), East Lansing, MI: Pearson, 2001.